\documentclass[aps,prd,preprint,superscriptaddress,tightenlines,nofootinbib,showpacs]{revtex4}
 
\usepackage{graphicx}% Include figure files
\usepackage{dcolumn}% Align table columns on decimal point
\usepackage{bm}% bold math

\begin{document}
\preprint{CLNS 03-1818}       % for CLNS notes
\preprint{CLEO 03-04}         % for CLNS notes

\title{Branching Fractions of $\tau$ Leptons to Three Charged Hadrons}
\author{R.~A.~Briere}
\author{G.~P.~Chen}
\author{T.~Ferguson}
\author{G.~Tatishvili}
\author{H.~Vogel}
\affiliation{Carnegie Mellon University, Pittsburgh, Pennsylvania 15213}
% ...
%\author{(CLEO Collaboration)} %FOR PRD_SPECIAL_CHANGEME
\collaboration{CLEO Collaboration} %FOR PRL,CLNS
\noaffiliation
\date{\today}  
\begin{abstract}
%bkh Based on a data sample of 3.26 fb$^{-1}$ collected with the  CLEO III detector,
%bkh we %study $\tau$ decays to three charged hadrons, and %
%bkh obtain 
From %bkh a dataset of  
%bkh 3.26 fb$^{-1}$ of 
electron-positron collision data
collected with the CLEO detector operating at CESR near
$\sqrt{s}$=10.6~GeV, 
improved measurements of the branching fractions for 
$\tau$ decays into three explicitly identified hadrons
and a neutrino are presented as 
${\cal B}(\tau^-\to \pi^-\pi^+\pi^-\nu_\tau)=(9.13\pm0.05\pm0.46)\%,$
${\cal B}(\tau^-\to K^-\pi^+\pi^-\nu_\tau)=(3.84\pm0.14\pm0.38)\times10^{-3},$ 
${\cal B}(\tau^-\to K^-K^+\pi^-\nu_\tau)=(1.55\pm0.06\pm0.09)\times10^{-3},$ 
and 
${\cal B}(\tau^-\to K^-K^+K^-\nu_\tau)<3.7\times10^{-5}$  
at 90\%~C.L., where the uncertainties 
are statistical and systematic, respectively.
%bkh , and remaining
%bkh $\pi^+\pi^-$ pairs from $K^0_S$ decay in the final state are
%bkh considered to background. 
\end{abstract}
\maketitle
  
%\section{Introduction} 
For hadronic $\tau$ decays, final states with kaons 
provide a powerful probe of the strange sector of the 
weak charged current.
In this Letter, we present the improved measurements of branching fractions 
for $\tau^-\to \pi^-\pi^+\pi^-\nu_\tau$, $K^-\pi^+\pi^-\nu_\tau$,  
and $ K^-K^+\pi^-\nu_\tau$ decays, %without $K_S^0$ excluded  
and an improved upper limit for the phase space suppressed and 
Cabibbo-suppressed decay $\tau^-\to K^-K^+K^-\nu_\tau$. 
(Charge conjugate decays are implied).
{%\bf  
The decay mode $\tau^-\to K^-\pi^+\pi^-\nu_\tau$ has a significant 
contribution to the overall strange spectral function 
which can provide a direct determination of the strange quark mass and, 
potentially, precision extraction  of the CKM element $V_{us}$~\cite{msvus}. 
%two fundamental parameters of the Standard Model. 
The decay $\tau^-\to K^-K^+\pi^-\nu_\tau$ can proceed via both the
vector and axial-vector currents~\cite{prd47} and, 
therefore, has a sensitivity to the Wess-Zumino anomaly~\cite{WZ}.  
}

The data used for this analysis were collected with the CLEO~III 
detector~\cite{CLEOIII} located at the symmetric $e^+e^-$ 
Cornell Electron Storage Ring (CESR).  
The data sample consists of 3.26~fb$^{-1}$ taken 
on or near the $\Upsilon(4S)$, % with a sysmmetric beam collider, 
corresponding to 2.97$\times10^6$ 
$\tau^+\tau^-$ pairs. The CLEO~III detector configuration %bkh
%bkh is an upgraded version of the CLEO II detector 
features a new four-layer silicon strip 
vertex detector, a new wire drift chamber, and,
most importantly for this analysis, a ring 
imaging Cherenkov (RICH) particle identification system. 
%bkh The tracking system consists of the silicon detector and the 
Particle trajectories, momenta, and charges are measured with the 
tracking system, which uses hits from both the silicon
detector and drift chamber. The specific ionization, $dE/dx$, measured
in the drift chamber's 16 inner axial and 31 outer stereo
layers is sensitive to the traversing particle's mass.  
The RICH detector~\cite{tom} surrounds the drift chamber, and uses 1~cm-thick 
LiF radiators to generate Cherenkov
photons from the incident charged particles. These photons then propagate 
%bkh in a 20-cm thick $N_2$ filled transparent 
through a 20~cm-thick expansion volume of gaseous nitrogen at atmospheric
pressure, and are 
%bkh where their cones of Cherenkov light 
%bkh expand and
%bkh The Cherenkov photons are then 
detected by multi-wire proportional chambers filled 
with a mixture of triethyleamine (TEA) and CH$_4$ gases,
and localized by readout from 8~mm-square cathode pads
located on a cylinder of $\sim$1~m radius.
RICH particle identification is available over the central region 
of the detector, $|\cos\theta|<0.83$, where $\theta$ is the polar angle 
with respect to the direction of the incident positron beam. 
{A detailed description of the RICH performance 
can be found in Ref.~\cite{RICH}.} 
The electromagnetic calorimeter surrounds the RICH detector
and measures the energy, position,
and lateral shape of showers induced by charged and neutral
particles. It contains 7784  
16-radiation-length-long CsI(Tl) crystals, arranged in a barrel section
($\vert\cos\theta\vert< 0.83$) and two endcaps 
($0.83<\vert\cos\theta\vert<0.95$).
%Photon and electron identification combines information from
%both the tracking system and calorimeter.
These components all operate inside a
superconducting solenoid coil which creates a
uniform magnetic field of 1.5~T. 
A muon detection system %bkh  located outside 
surrounds the solenoid coil %bkh which consists of
with 1~m of iron absorber interspersed with Iarocci tube
wire chambers operated in proportional mode. 
%bkh Muons above
%bkh 1~GeV/c can be detected over $|\cos\theta|<0.85$.  A 
%bkh typical muon efficiency in the barrel region of the detector
%bkh is $\sim90\%$, with 2-5\% probability for hadrons to punch through
%bkh to fake a penetrating muon.

%bkh  Previously, the most precise measurements of branching fractions  for
%bkh $\tau$  decays to three charged hadrons~\cite{cleoK2pi,opalK2pi,alephK2pi}  
%bkh were all based on the particle separation provided 
%bkh by ionization energy loss ($dE/dx$) in wire-gas tracking chambers. 
%bkh The enhanced charged particle identification
%bkh provided by the CLEO RICH system complements the continued
%bkh high performance level of the entire CLEO detector and available 
%bkh luminosity from CESR to result in the substantial improvements
%bkh described here.

We combine RICH and $dE/dx$ information to determine if a
track appears more consistent with a pion or kaon identity.
The %bkh standard 
deviation of the measured $dE/dx$ for any track 
from that expected under particle hypothesis $i$ 
($i=e$, $\mu$, $\pi$, $K$, $p$) at that momentum, 
in units of the expected Gaussian width of the distribution, is
defined as $\delta_i$. The $dE/dx$ resolution is about 6\% for hadrons.   
For each detected charged particle track, the RICH detector response is
condensed into a $\chi^2_i$-like variable %~\cite{RICH} 
% $\chi^2_i\equiv -2\ln{\cal L}_i$ from the likelihood ${\cal L}_i$
for each particle hypothesis. The value of  $\chi^2_i$ 
is derived from %Gaussian-like probability function ${G}$ of 
the number $N_\gamma^{i}$ of detected Cherenkov photons
 and their locations relative to the Cherenkov cone expected 
for a particle with that momentum and mass.  
%$${\cal L}_i=\prod\limits_{j=1}^{N_\gamma^{obs}}
%\left[{G}(\theta_j^{obs}|\theta_i^{exp})+B\right].$$ 
%Where $G$ is a Gaussian-like function expressing the probability of observing 
%the $j$th photon at an angle $\theta_j^{obs}$ relative to the 
%expected Cherenkov angle $\theta_i^{exp}$ for a particle with 
%that momentum and mass. 
%where $B$ is the background probability approximately by a flat function of 
%$\theta_j^{obs}-\theta_i^{exp}$.   
%The average number of photons detected is twelve, the 
%Cherenkov angle resolution is 13 mrad.  
%The performance of the RICH detector is described 
%elsewhere~\cite{RICH} in detail.
%bkh The difference between these likelihoods for particle hypotheses $i$ and $j$,
%bkh defined  by  $\chi^2_i-\chi^2_j=-2\ln(L_i/L_j$),
%bkh To get better particle separation,
Pions and kaons are distinguished from one another 
with the combined RICH-$dE/dx$ variable
$\Delta\chi^2=\chi^2_\pi-\chi^2_K+\delta_\pi^2-\delta_K^2$.
We identify a track as a pion (kaon) if 
$\vert\cos\theta\vert< 0.83$, $N_\gamma^\pi(N_\gamma^K)\ge 3$,
 $\Delta\chi^2 < (>)0$, and apply these criteria to
three-prong-side tracks. For the pion in
$\tau^-\to K^-K^+\pi^-\nu_\tau$, $|\delta_\pi|<3.0$ 
alone is used, yielding $\sim$10\% higher efficiency from 
a looser criterion as well as the region
outside the RICH acceptance,
$0.83 < \vert\cos\theta\vert < 0.93$. The consequently higher
$K$-fakes-$\pi$ rate is unimportant for this mode due to
the dearth of three-kaon events.

%bkh In addition, a cut on minimum number of Cherenkov photons per 
%bkh track is imposed. 

\begin{figure}[hbtp]
\includegraphics*[width=0.9\textwidth]{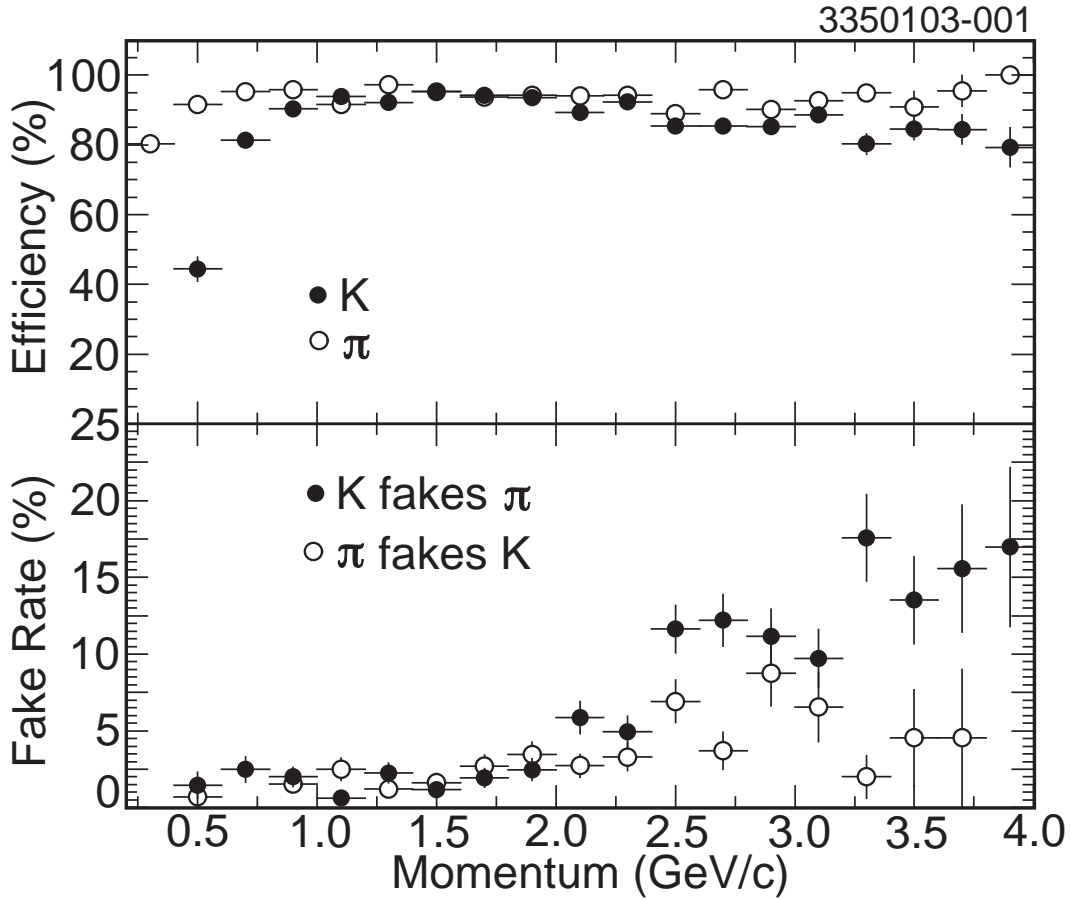} 
\caption{Kaon (solid circles) and pion (open circles) identification 
efficiencies (top) and 
fake rates (bottom) as a function of their momenta for tracks already
within $\vert \cos\theta\vert < 0.83$ as determined from
$D^0\to K^-\pi^+$ decays. Error bars 
are statistical only.} 
\label{effs}
\end{figure}

We take advantage of the copious cascade decays $D^{*+}\to D^0\pi^+$ 
with $D^0\to K^-\pi^+$, in which the
charged kaon and pion can be tagged kinematically without
reference to RICH or $dE/dx$ information, to measure 
the probability for pions and kaons to be correctly or incorrectly
identified. This method measures the area above background
of the $K^-\pi^+$ mass peak with and without the particle identification
%bkh $\Delta\chi^2$ 
criteria to determine the efficiency and fake rates for
both pions and kaons. We find that the efficiency for
pion and kaon identification ranges from 80$-$95\% over most of the momentum
range of interest for this analysis (0.5$-$3~GeV/$c$), and that
the probability for a pion to fake a kaon or vice versa
ranges from 1$-$2\% for momentum values of 0.5$-$2~GeV/$c$ 
and rises to about 10$-$15\%
for momenta around 3~GeV/$c$, as shown in Fig.~\ref{effs}.

%\section{Event Selection} 

Events are required 
to have four well-measured charged particles 
in a one-versus-three topology (the two hemispheres being defined
by the event thrust axis)
%(a lone track more than 90$^\circ$
%from the other three)
%bkh passing the track quality requirements with 
and zero net charge.   
We use 1-prong $\tau$ leptonic ($e/\mu$) or hadronic 
decays ($\rho/\pi$) to ``tag'' the event. %bkh the other $\tau$. 
%bkh $^-\to K^+K^-\pi^-\nu_\tau$. 
%An electron candidate is selected based on $dE/dx$ information and 
%the ratio of the associated shower energy in the calorimeter to the 
%measured track momentum. 
%bkhAn electron candidate is identified by its specific ionization 
%being higher than two standard deviations below the most probable value for
%an electron at that momentum  i.e. 
%bkh
An electron tag requires the lone track to have
$\delta_e>-2$ and $E/p$, the ratio of energy deposited in the calorimeter to 
track momentum, to satisfy $0.85 < E/p < 1.1$. 
These criteria attain $\sim95\%$ efficiency and 
a pion-faking-electron probability less than 1\%.
%bkh of pions will fake an electron.
A muon tag must have the lone track penetrate 
at least  three (if $p<$2~GeV/$c$) or five (if $p>$2~GeV/$c$) absorption 
lengths of material, which achieves 80$-$90\% efficiency
and a few percent pion-faking-muon probability for $p>$1~GeV/$c$.

If the tag-side track 
is not identified as an electron or a muon,
tag-side calorimeter showers are then examined.
%bkh The events that do not satisfy the lepton criteria are classified as 
%bkh hadron tags. 
If the invariant mass of the lone track
and showers in the tagging hemisphere not matched to that track 
is below 0.5~GeV/$c^2$, the event is classified as $\pi$-tag. 
To establish a $\rho^-(\to\pi^-\pi^0)$-tag, we require a 
$\pi^0\to\gamma\gamma$ candidate on the 1-prong side.
For photon candidates, %bkh  from $\pi^0$ decays, 
%bkh unmatched 
we select showers which do not match the projection of any charged track
 trajectories %with $|\cos\theta|<0.95$ 
into the calorimeter,
have energies exceeding 30~(50)~MeV in
%bkh good section of the 
%barrel or 50 MeV in the endcap and to be consistent with 
the barrel (endcaps), and have compact
lateral profiles consistent with 
%bkh the shapes 
those expected from electromagnetic showers. 
%bkh To  get better $\pi^0$ mass resolution and signal-to-background, 
%bkh we require that 
At least one photon from any $\pi^0$ candidate decay
must be located in the barrel.
%bkh of the calorimeter. %good section of the barrel. 
 We take the most energetic $\gamma\gamma$ pair which has
invariant mass within
$(-3.0,2.5)$ standard deviations of $m_{\pi^0}$ (typically
a window $\sim$30~MeV/$c^2$ wide).
After kinematically constraining the $\gamma\gamma$ pair to $m_{\pi^0}$,
the invariant mass of this fitted $\pi^0$ candidate
and the associated charged track, assigned the pion mass, is required
to be consistent with 
the $\rho$ mass: $0.5< m(\pi^-\pi^0) < 1.0$~GeV/$c^2$. 
%bkh All other 

%bkh A typical signal event should not deposit any significant 
%bkh extra energy in the calorimeter. In most cases, extra
%bkh unmatched 
%bkh energy deposited in the calorimeter is a signature
%bkh of various backgrounds with one or more $\pi^0$'s. To 
To suppress 
%bkh these 
%bkh tag-side and signal-side 
backgrounds with one or more neutral particles, %bkh $\pi^0$'s, 
we do not allow 
any shower in the event unassociated with a charged track or a $\pi^0$
participating in $\rho$-tagging  
with energy exceeding 100~MeV. 
After the event selection, the backgrounds with extra neutrals 
 contribute 4.1\%, 2.1\% and  2.6\% of the total observed events 
for the channels with 0, 1 and 2 kaons, respectively. 
%{ The remaining backgrounds %relative to
%the signal events are suppressed to below 4\%. 
%4.3\%, 3.8\% and 3.0\% for the decays
%$\tau^-\to\pi^+\pi^-\pi^-\nu_\tau$, $K^-\pi^+\pi^-\nu_\tau$ and 
%$K^+K^-\pi^-\nu_\tau$ respectively.  } 
%bkh $E_\gamma>0.1$ GeV. 
%in the tagging hemisphere or in the signal hemisphere. 
In order to reject 
%bkh the 
two-photon backgrounds (characterized by the 
missing momentum along the beam direction and low visible energy),
%unlike tau-pair events which have neutrinos carrying away
%undetected energy in no preferred direction),
%bkh we require the absolute value of the cosine of the missing 
%momentum  to be less than 0.95 %,
the missing momentum must point into the CLEO detector 
($|\cos\theta_{miss}|<0.95$)
and the visible 
energy must exceed 20\% of the $e^+e^-$ center-of-mass energy.
%Additional background suppression for the $K^-K^+\pi^-$ mode  
%is obtained by requiring the net three-particle momentum
%component transverse to the beam line to exceed 2~GeV/$c$. 
%which is decay modes dependent. %$E_{vis}/E_{cm}>0.4$.
For all modes, we further demand that the three-prong mass, calculated
using mass assignments determined from particle identification, lie below the 
$\tau$ mass. For the decay $\tau^-\to K^-\pi^+\pi^-\nu_\tau$, a 
veto of $K_S^0$ candidates with $K_S^0\to\pi^+\pi^-$ within 10~MeV/$c^2$ of 
the $K_S^0$ nominal mass and a production vertex %bkh at a distance of 
more than 1~cm from the primary interaction point is applied. 
Any $K_s^0$ content in both modes with two or more pions
is treated as background, not signal.

%bkh Based on the generic $\tau$ and continuum MC samples   
%bkh   Based on the generic $\tau$ and continuum MC samples   
%bkh generated by KORALB~\cite{korb} and  JETSET~\cite{jetset}
%bkh  and simulated with GEANT~\cite{GEANT},
%bkh we optimize   
%bkh the particle identification cuts $\Delta \chi^2$
%=\chi^2_{\pi}-\chi^2_{K}+ \sigma_\pi^2-\sigma_K^2$ 
%bkh and the numbers of 
%bkh Cherenkov photons $N_\gamma^K$, $N_\gamma^\pi$ to
%bkh maximize  signal significance 
%bkh which  are determined to be  
%bkh $N_\gamma^K\ge3$, $\Delta \chi^2>0$ for kaons; and $N_\gamma^\pi\ge3$, 
%bkh $\Delta \chi^2<0$ for pions. 
%bkh For $\tau^-\to K^+K^-\pi^-\nu_\tau$, $dE/dx$
%bkh is used to identify the pion with a requirement $|\sigma_\pi|<3.0$ 
%bkh to get about 10\% more efficiency; for other modes, all particles are
%bkh identified using the $\Delta \chi^2$ and $N_\gamma$ requirements.   

%\subsection{The Backgrounds} 

 Efficiencies and the remaining backgrounds are evaluated 
with Monte Carlo (MC) events
from the KORALB~\cite{korb} ($\tau$-pairs) and  JETSET~\cite{jetset}
($q\bar{q}\to$~hadrons) generators
passed through the GEANT-based~\cite{GEANT} CLEO detector simulation.
Any MC event which has a tag-side particle which is
truly  $e^\pm$, $\mu^\pm$, $\pi^\pm$, $K^\pm$, or $\rho^\pm$ is considered
signal for purposes of efficiency; other tags are considered
background even if the three-prong side is a signal decay mode.
For $\tau^-\to K^-K^+K^-\nu_\tau$ decay, we generate signal MC 
samples according to phase space. 
The particle identification simulation
models the general features of the efficiency and fake
rates reasonably well (at the few percent level). 
We correct the 
Monte Carlo efficiencies and fake rates by hand,
using momentum-dependent scale factors derived from
measured and MC particle identification rates. %, thus 
%using the measured rates from data instead of those in the 
%making use of the measured rates from the data instead of those in the 
%Monte Carlo simulation.
%bkh After the event selection, two photon backgrounds are negligible.
%bkh We use the simulated generic $\tau$ and continuum MC samples 
%bkh to study the backgrounds. 
%With RICH, 

The corrected efficiencies for events in which one $\tau$
decays into a tag mode and the other into the relevant signal mode
(integrated over all tag modes, which constitute about 72.4\% 
of all $\tau$~decays), 
observed numbers of events in the data, 
and the expected numbers of  $\tau$ and  $q\bar{q}$~background 
events  are given in Table~\ref{table1}. There are
substantial signals evident in all but the three-kaon mode.
%bkh Backgrounds from $q\bar{q}$ are suppressed to less than the 
%bkh level of 2\%.
The largest backgrounds come from $\tau$ decays to three charged hadrons 
%bkh with or without neutrals 
with a single misidentification of a pion
or kaon. Many such events are eliminated by the restriction
of three-prong mass to be below the $\tau$ mass. 
%bkh The $\tau^-\to\pi^+\pi^-\pi^-\nu_\tau$ decay contributes
Each decay in Table~\ref{table1} contributes 
the dominant background for the channel 
below it due to the particle misidentification.
For the $\tau^-\to K^-K^+K^-\nu_\tau$ decay,  %bkh with 
the $\tau$~background is
not taken from the Monte Carlo sample, but rather
the measured $\pi$-fakes-$K$ rate is applied to the 
$K^-K^+\pi^-$ events in the data (many of which subsequently fail
the three-prong mass restriction).

\begin{figure}[hbtp]
\includegraphics*[width=0.9\textwidth]{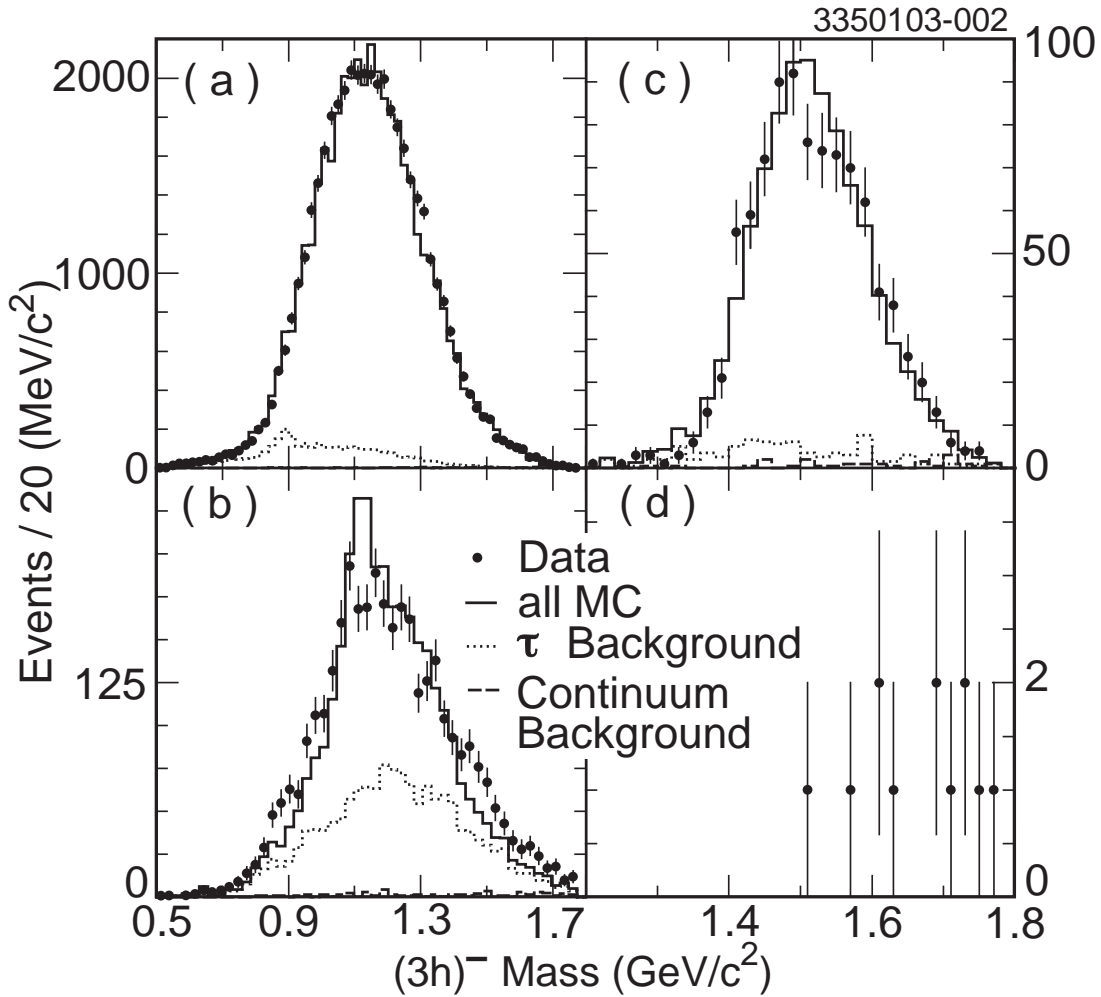} 
\caption{Three-hadron masses from data (solid circles)
and MC, $\tau$-pair 
background (short dashes), $q\bar{q}$ background (long dashes),
sum of background and signal (solid lines), for
$(a)$ $\tau^-\to\pi^-\pi^+\pi^-\nu_\tau$,
$(b)$ $\tau^-\to K^-\pi^+\pi^-\nu_\tau$, 
$(c)$ $\tau^-\to K^-K^+\pi^-\nu_\tau$ 
and 
$(d)$ $\tau^-\to K^-K^+K^-\nu_\tau$. 
} 
\label{fig01}
\end{figure}

\begin{figure}[hbtp]
\includegraphics*[width=0.9\textwidth]{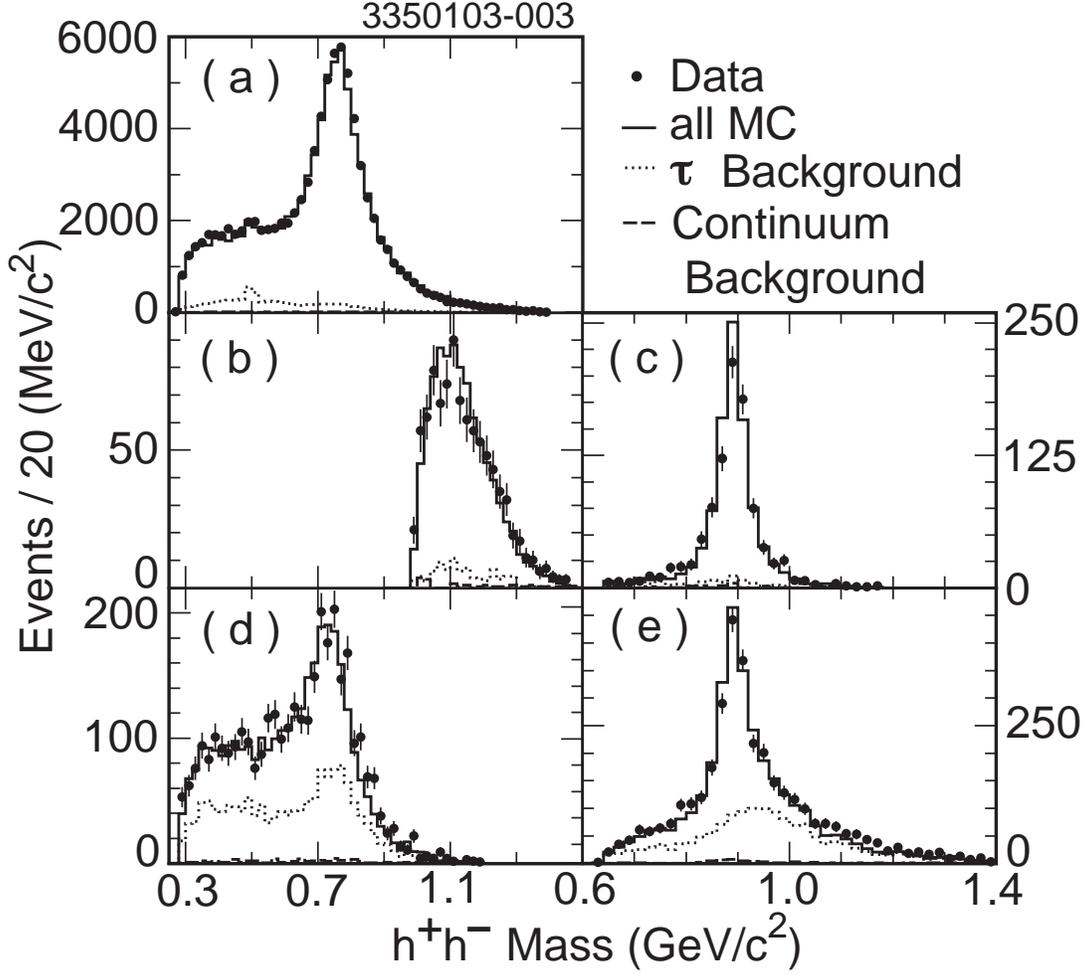} 
\caption{Mass distributions for two oppositely charged hadrons 
showing the data (solid circles)
and MC [$\tau$-pair 
background (short dashes), $q\bar{q}$ background (long dashes),
sum of background and signal (solid lines)]
for $(a)$ $\tau^-\to\pi^-\pi^+\pi^-\nu_\tau$ (two
entries per event),  $(b)$ and $(c)$
$\tau^-\to K^-K^+\pi^-\nu_\tau$,
 ($K^+K^-$ and $K^+\pi^-$ masses, respectively) 
 and  $(d)$ and $(e)$ $\tau^-\to K^-\pi^+\pi^-\nu_\tau$ 
 ($\pi^+\pi^-$ and $K^-\pi^+$ masses, respectively).} 
\label{fig02}
\end{figure}

%\section{Results} 

%bkh To determine the detection efficiencies, we generate signal MC samples
%bkh using KORALB~\cite{korb,mytune}
%bkh and  simulate the CLEO detector response 
%bkh with GEANT in the same manner as the data. 
%bkh The overall detection efficiencies %bkh after the event selection
%bkh integrated over all tags are given in 
%bkh Table~\ref{table1}. 

In Fig.~\ref{fig01}, we show the three-hadron masses
for the four modes considered here along with the 
results from the overlaid MC simulation.
%bkh comparison of the hadronic 
%bkh masses for three hadrons after the event selection. 
The two-particle substructure is presented 
in Fig.~\ref{fig02}. 
Good agreement between the data and MC is observed~\cite{mytune,pi2pi0}.
% The 
%CLEO results~\cite{mytune,pi2pi0} on the structures of the 
%decays $\tau^-\to (3\pi)^-\nu_\tau$
%and $\tau^-\to K^-\pi^+\pi^-\nu_\tau$ have been implemented 
%in the KORALB~\cite{korb}. %For the decay 
%The three-pion subsample is dominated 
%by $\tau^-\to\rho^0\pi^-\nu_\tau$ production~\cite{pi2pi0}. 
%The decay $\tau^-\to K^{*0}K^-\nu_\tau$ 
%dominates the $K^+K^-\pi^-$ events. 
The $\tau^-\to K^-K^+\pi^-\nu_\tau$ mode can be used to measure the 
contribution from the Wess-Zumino anomaly, offering 
an alternative channel to % other than the decay 
$\tau^-\to\eta\pi^-\pi^0\nu_\tau$. 
A quantitative estimate requires determination of 
the structure functions~\cite{prd47}, for which a substantially larger
data sample would be needed. %bkh  will be necessary for such an analysis. 
%The $K^-\pi^+\pi^-$ final state results from the mixing of 
%the $K_1(1270)$ and $K_1(1400)$ which subsequently decay into a $\rho$ or  
%$\overline{K^{*0}}$ and a pion~\cite{mixing}. 

{
\begin{table}[hbtp]%{ 0.45\textwidth}
\caption{%Measurements of $\tau$ three prong decays with  statistical and 
%systematic uncertainties, respectively and upper limit at CL=90\%.} 
Candidate event yields for the data, estimated $\tau$ and $q\bar{q}$ %continuum
background event totals, and overall efficiencies $\epsilon$ 
%and total systematic errors 
for $\tau$ three-prong decays.} 

\begin{center}
\begin{tabular}{lccccc} \hline
 \ \ \ Mode & data  &  $\tau$~bgd &  $q\bar{q}$~bgd & $\epsilon$(\%) \\ \hline
%& & background & background & & \\  \hline 
$\pi^-\pi^+\pi^-\nu_\tau$ & 43543 & 3207$\pm$57  & 152$\pm$12  & 
 10.27$\pm$0.08  \\ %\hline 
$K^-\pi^+\pi^-\nu_\tau$ & 3454 & 1475$\pm$38 & 57$\pm$8  & 
  11.63$\pm0.12$  & \\ %\hline 
$K^-K^+\pi^-\nu_\tau$ & 932   & 86$\pm$9 & 19$\pm$4   & 
12.48$\pm$0.11 \\ %\hline  
$K^-K^+K^-\nu_\tau$ & 12 & 4$\pm$2 & $0.4\pm0.6$   & 
9.43$\pm0.10$ \\ \hline  
\end{tabular}
\end{center}
\label{table1}
\end{table}
}

%\section{Systematics} 
\label{systematics4}
%bkh The dominant contribution to the systematic errors  
%bkh comes from the event selection efficiencies. 
%bkh The uncertainty arising from the $E_\gamma<0.1$ GeV requirement for
%bkh the extra shower is 3.0\% for $\tau^-\to K^+K^-K^-\nu_\tau$ decay and 
%bkh less than 1\% for the other decays. The systematic error 
%bkh related to model dependent efficiency is less than 2.0\%. 
%bkh The background subtraction contributes a 
%bkh systematic uncertainty of 7.8\% for the $\tau^-\to K^+K^-K^-\nu_\tau$ decay
%bkh and less than 2.5\% for the others which 
%bkh includes uncertainties from the $\tau$ and continuum background subtraction %,  
%bkh  and from the particle identification fake rates. 
%The backgrounds due to the fake rates are estimated  
%by reweighting the momentum dependent fake rates from the data, the 
%difference between the reweighted backgrounds and those from  
%MC is taken as a systematic error.  
%bkh The systematic error from the cut on the number of tracks
%bkh in an event is about 2.0\%. The tracking efficiency uncertainty 
%bkh is 0.5\% per track. 
%bkh The uncertainty of particle identification efficiency is 1.5\% per 
%bkh kaon and 0.5\% per pion.  Also included in the systematic error
%bkh are uncertainties in the number of $\tau^+\tau^-$ pairs (2.2\%) 
%bkh and MC statistics (1.0\%). All these contributions are added in quadrature.
%bkh The total systematic errors are 
%bkh  4.4\%, 5.1\%, 5.9\% and 10.2\% for the decays 
%bkh listed in Table~\ref{table1}, respectively. 

  Most sources of systematic uncertainty affect the four
modes similarly. MC statistics for signal events
add 1\% uncertainty. The uncertainty in the number of $\tau$-pair 
events produced, and therefore in the normalization of
the branching fractions,
derives from the uncertainty of luminosity~\cite{lumins} (2\%) and
 $\tau$-pair cross section (2\%)~\cite{korb}. 
Studies comparing track-finding efficiencies in the MC simulation to data
result in a 0.5\% per track uncertainty, added linearly to 
become 2\% for the entire event.
Several effects are probed individually
by tightening and/or relaxing an event selection criterion,
recalculating the branching fractions, and assigning
an uncertainty in how well the MC models reality 
based on the changes induced: 
% which include the  uncertainties %include %an extra
%track incorrectly found in signal events (1.7\%),
allowing for an extraneously reconstructed track in signal events adds 2\% 
uncertainty; 
one systematic error arising from the assumed absence of 
two-photon physics background %(0.5\%)
is 0.5\% which can be tested by
imposing a harsh net transverse momentum of the
three-prong side requirement of 3~GeV/$c$;  
uncertainty due to the background levels from decay modes
 with extra neutrals
allowed in is estimated by relaxing the veto energy on extra calorimeters
showers to as high as 250~MeV and found to be 1\%; and 
 variations
of the remaining non-particle-identification 
selection criteria contribute an error of 1\%. Uncertainty
in $\tau$ branching fractions results in 1\%
error due to changes in $\tau$ backgrounds.
The efficiencies of particle identification
were studied exhaustively for variations with
time (0.5\%) or charged-track multiplicity (0.5\%),
bias of the efficiency-measuring technique (0.5\%), the data/MC 
efficiency correction procedure (0.4\%),
and statistics of the $D^*$ data samples (0.3\%), which
all sum in quadrature to 1\% per pion or kaon, which then are added 
linearly for the event as a whole to 3\% overall.
When the branching fractions are broken out by
individual tags, consistent results are obtained,
verifying both the modeling of tag efficiencies and
the lack of unexpected non-$\tau$ backgrounds which sometimes
show up in the non-leptonic tags.
Together these errors sum in quadrature to 5\%, which
applies to the branching fractions of all four modes.

  Other systematics vary from mode to mode, which will
be referred to in the order they appear in Table~\ref{table1}. 
To estimate uncertainties from $q\bar{q}$ backgrounds, we compare
event totals above the $\tau$ mass % with those predicted
between the data and $q\bar{q}$ MC,  
%result in an uncertainty assignment of $\sim$50\%
and assign an additional $\sim$50\% (relative) uncertainty 
from the observed differences. This %in the  $q\bar{q}$ background, 
results in uncertainties of
0.2\%, 2\%, 1\%, and 3\% for the four modes, respectively. 
An overall scale 
uncertainty in the fake rate per pion or kaon of 0.25\% (absolute) %, 
%or 10$-$25\% (relative)
%over the momentum range of interest, 
is estimated by comparing
measured results to those predicted by the RICH simulation,
cross-checking with MC for possible bias in the fake-rate
measurement technique, 
examining the rate of the forbidden $\tau^-\to K^+\pi^-\pi^-\nu_\tau$
(which is dominated by $\tau^-\to \pi^+\pi^-\pi^-\nu_\tau$ with
a pion faking a kaon),
and estimating the error from the momentum-reweighting correction
procedure. This fake rate uncertainty propagates to the
branching fractions differently because of different
levels of feed-down: 0.1\%, 9\%, 2\%, and 12\%, respectively.
The $K^-K^+\pi^-$ mode substructure is not well understood, 
and variations in the possibilities
result in a 2\% uncertainty in the efficiency.
The three kaon mode substructure is 
completely unknown and assumed to occur via phase space
with no assigned error. The remaining modes appear to
be adequately described by the models used. Adding these contributions
together in quadrature with the common sources results in
total relative systematic uncertainties of  5\%, 10\%, 6\%, and 14\%,
respectively.

%bkh After the backrounds have been subtracted, we determine the 
%bkh branching fractions.  
For all decay modes, the yields are obtained
from the numbers of the observed events with the backgrounds 
subtracted, %After the yields corrected by the detection efficiencies 
%and the number of the produced $\tau$ pairs,  
the measured branching fractions (with $K_S^0$ excluded from
the modes with two or more pions) are
%bkh and an upper limit for $\tau^-\to K^-K^+K^-\nu_\tau$ decay at
%bkh 90\% confidence level are  summarized as follows: 
\begin{eqnarray}
{\cal B}(\tau^-\to \pi^-\pi^+\pi^-\nu_\tau)&=&(9.13\pm0.05\pm0.46)\%,\nonumber\\
{\cal B}(\tau^-\to K^-\pi^+\pi^-\nu_\tau)&=&(3.84\pm0.14\pm0.38)\times10^{-3},\nonumber \\ 
{\cal B}(\tau^-\to K^-K^+\pi^-\nu_\tau)&=&(1.55\pm0.06\pm0.09)\times10^{-3},
\nonumber\\  
{\cal B}(\tau^-\to K^-K^+K^-\nu_\tau)&<&3.7\times10^{-5} {\rm \ \ at\ \
90\%\ CL}.\nonumber
\end{eqnarray}
The errors are statistical and systematic, respectively. 
For the decay $\tau^-\to K^-K^+K^-\nu_\tau$, %the yield is obtained 
%from the observed events with the backgrounds subtracted, 
the upper limit is obtained according to the procedure 
described in the PDG~\cite{PDG} %after the backgrounds subtracted 
with both statistical and systematic errors taken into accout.  
%For the decay $\tau^-\to K^-K^+K^-\nu_\tau$, the number of the 
%observed events is comparable with the backgrounds within three
%standard deviations and we choose to provide an upper limit 
%for the corresponding branching fraction using the background 
%subtracted number of events, and the upper limit is
%\begin{eqnarray}
%{\cal B}(\tau^-\to K^-K^+K^-\nu_\tau)&<&3.7\times10^{-5} {\rm \ \ at\ \
%90\%\ CL}.\nonumber
%\end{eqnarray}
%The upper limit at CL=90\%  is taken as 1.64 standard deviation 
%away from the central value~\cite{PDG} after the backgrounds subtracted 
%according to the procedure described in PDG~\cite{PDG} 
%with both the statistical and systematic errors taken into account. 
This represents the first direct measurement of 
${\cal B}(\tau^-\to \pi^-\pi^+\pi^-\nu_\tau)$ with three pions 
explicitly identified, and our result 
is in good agreement with the PDG fitted value %bkh $(9.22\pm0.10)\%$
derived from the PDG constrained fit to the measurements of 
$\tau^-\to (3h)^-\nu_\tau$ decays with and without identified
charged kaons~\cite{PDG}.
%The measurements of  ${\cal B}(\tau^-\to K^-K^+\pi^-\nu_\tau)$ 
%and ${\cal B}(\tau^-\to K^-\pi^+\pi^-\nu_\tau)$ 
%represent factors of two or more improvement in precision over
%previous measurements~\cite{PDG}, with which these
%central values are consistent.
 The measurement of ${\cal B}(\tau^-\to K^-K^+\pi^-\nu_\tau)$ 
is consistent with the previous value~\cite{PDG} and improves upon the
precision previously attained by a factor of two. 
%bkh agrees well with the PDG fitted result %bkh $(1.60\pm0.19)\times10^{-3}$
%bkh but has half the uncertainty. 
%bkh Our result for 
%bkh is consistent with 
%bkh the results %bkh $(3.46\pm0.23\pm0.56)\times10^{-3}$ 
Our result ${\cal B}(\tau^-\to K^-\pi^+\pi^-\nu_\tau)$ 
is more precise than and 
consistent with both the CLEO~\cite{cleoK2pi} 
and %bkh $(3.60\pm0.82\pm0.48)\times10^{-3}$ from
OPAL~\cite{opalK2pi} results, 
%bkh but significantly higher than the %bkh result 
%bkh $(2.14\pm0.37\pm0.29)\times10^{-3}$ from 
while it is larger than the ALEPH~\cite{alephK2pi} value by
nearly three standard deviations. %, and has a smaller uncertainty. 
%and taking both the statistical and systematics errors into account. 
%\begin{eqnarray}
%{\cal B}(\tau^-\to K^-K^+K^-\nu_\tau)<3.7\times10^{-5} {\rm \ \ at\ \
%90\%\ CL}.\nonumber
%\end{eqnarray}
The upper limit for the Cabibbo-suppressed decay
 $\tau^-\to K^-K^+K^-\nu_\tau$ has been improved over
the PDG value~\cite{PDG} by a factor of five. 
%bkh ${\cal B}(\tau^-\to K^-K^+K^-\nu_\tau)<1.9\times10^{-4}$
%bkh Taking advantage of the good particle separation 
%bkh capability at CLEO III, we have improved the measurements of 
%bkh $\tau$ decays to three charged hadrons. 
Taken together, these measurements significantly enhance
our understanding of the kaon content of three-prong $\tau$ decays.

%\acknowledgements

We gratefully acknowledge the effort of the CESR staff 
in providing us with
excellent luminosity and running conditions.
This work was supported by 
the National Science Foundation,
the U.S. Department of Energy,
the Research Corporation,
and the 
Texas Advanced Research Program.


\begin{thebibliography}{99}

\bibitem{msvus}{ALEPH Collaboration, R. Barate {\it et al.}, 
Eur. Phys. J. C {\bf 11}, 599 (1999); S. Chen 
{\it et al.}, Eur. Phys. J. C {\bf 22}, 31 (2001); 
A. Pich and J. Prades, JHEP {\bf 10}, 004 (1999); S. Narison, 
Phys. Lett. {\bf B466}, 345 (1999). }       

\bibitem{prd47}R. Decker {\it et al.}, Phys. Rev. D {\bf 47}, 4012 (1993). 

\bibitem{WZ}J. Wess and B. Zumino, Phys. Lett. {\bf B37}, 95 (1971). 

%\bibitem{alphastau}{\bf CLEO Collaboration, T. Coan {\it et al.}, 
%Phys. Lett. {\bf B356}, 580 (1995); ALEPH Collaboration, R. Barate 
%{\it et al.}, Eur. Phys. J. C {\bf 4}, 409 (1998). }  

\bibitem{CLEOIII} CLEO Collaboration, Y. Kubota {\it et al.}, Nucl. Instrum.
                  Methods  A {\bf 320}, 66 (1992); T.~Hill, Nucl. Instrum.
                  Methods  A {\bf 418}, 32 (1998); 
                  D. Peterson {\it et al.}, Nucl. Instrum. Methods A {\bf 478}, 
                  142 (2002). 

\bibitem{tom}T.  Coan,  Nucl. Instrum. Methods A {\bf 379}, 448 (1996).
\bibitem{RICH} M. Artuso {\it et al.}, hep-ex/0209009, talk presented at Fourth 
Workshop on RICH Detectors, Pylos, Greece, June 2002, to appear in 
the proceedings. %T.  Coan, NIM A379, 448 (1996).  
\bibitem{korb}S. Jadach and Z. Was, Comput. Phys. Commun. {\bf 36}, 191 (1985)
and {\it ibid} {\bf 64}, 267 (1991), {\it ibid} {\bf 76}, 361 (1993), 
{\it ibid} {\bf 85}, 453 (1995); R. Decker {\it et al.}, Z. Phys. 
C {\bf 58}, 445 (1993); M. Finkemeier and E. Mirkes, Z. Phys. 
C {\bf 69}, 243 (1996). 
The KORALB package~\cite{korb} does not include 
higher order contributions to the $\tau$ production cross section.
%The efficiency of the event selection is calculated using the same 
%MC, and therefore the error on the production cross 
%section and detection efficiency is small in comparison with 
%other systematic errors and is negligible in this analysis.
 We take the difference (2\%) between the cross sections from KORALB and 
its successor ${\cal KK}$ MC with higher order corrections 
as the systematics due to the $\tau$ pair cross section. 
For the ${\cal KK}$ MC, see S. Jadach, B. F. L. Ward 
and Z. Was, Comput. Phys. Commun. {\bf 130}, 260 (2000). 

\bibitem{jetset} T. Sjostrand, CERN-TH-6488-92 (unpublished).  

\bibitem{GEANT} R. Brun {\it et al.}, GEANT3.14, CERN DD/EE/84-1 (unpublished). 


\bibitem{mytune}For $\tau^-\to K^-K^+\pi^-\nu_\tau$ decay, we 
obtain the good data-MC agreement of mass distributions
shown solely by tuning the parameters of the form factor. 
We increase the $\rho'$ contribution
and suppress the $K^{*0}$ component on the low mass side 
 compared to the default model, switching off 
the $\rho''$ contribution and renormalizing it. %bkh Additional data 

\bibitem{pi2pi0} All CLEO results on the branching fractions and 
resonance contents of the decays $\tau^-\to (3\pi)^-\nu_\tau$ and 
$\tau^-\to K^-\pi^+\pi^-\nu_\tau$ have been implemented in the 
Monte Carlo program. For experimental 
results, see D. M. Asner {\it et al.}, Phys. Rev. D {\bf 61}, 012002 (2000),  
and D. M. Asner {\it et al.}, Phys. Rev. D {\bf  62}, 072006 (2000), 
see also A. Weinstein, 
hep-ex/0210058, talk given at Tau'02, Santa Cruz,
California, September 2002, to appear in the proceedings.

%\bibitem{mixing} CLEO Collaboration, D. A. Asner {\it et al.}, 

\bibitem{lumins} CLEO Collaboration, G.~Crawford {\it et al.},
Nucl. Instrum. Meth. A {\bf 345}, 429 (1994).


\bibitem{PDG} Particle Data Group, K. Hagiwara {\it et al.}, 
Phys. Rev. D {\bf 66}, 010001 (2002). 

\bibitem{cleoK2pi}CLEO Collaboration, S. Richichi 
{\it et al.}, Phys. Rev. D {\bf 60}, 112002 (1999). 

\bibitem{opalK2pi}OPAL Collaboration,  G. Abbiendi {\it et al.}, Eur.
 Phys. J. C {\bf 13}, 197 (2000). 

\bibitem{alephK2pi}ALEPH Collaboration, R. Barate {\it et al.}, Eur. 
Phys. J. C {\bf1}, 65 (1998).

\end{thebibliography}
\end{document}